\documentclass{aa}
\usepackage{psfig}

\begin{document}
%
   \title{Molecular gas and dust in NGC\,4550} 
   \subtitle{A galaxy with two counterrotating stellar disks}

   \author{T.~Wiklind\inst{1}
   \and
           C.~Henkel\inst{2}}
   \offprints{Wiklind, \email{tommy@oso.chalmers.se}}
   \institute{
              Onsala Space Observatory, S--43992 Onsala, Sweden
   \and
              Max--Planck--Institute f\"{u}r Radioastronomie,
              Auf dem H\"{u}gel 59, D--53121 Bonn, Germany
             }
   \date{Received 1 January 2001 / Accepted 1 January 2001}
%
\abstract{
We report the detection of $1 \times 10^7$\,M$_{\odot}$ of molecular 
gas in the central region of the S0/E7 galaxy NGC\,4550, inferred 
from observations of CO(1--0) emission. Dust is detected in HST WFPC2 
images and found to be asymmetrically distributed around the nucleus, 
only extending to a galactocentric distance of 7$''$ (600 pc). The 
shape of the CO emission profile is consistent with a molecular gas 
distribution following the dust. The distribution of the dust and 
gas in the center could be the result of an $m=1$ instability, which 
is the fastest growing unstable mode in counterrotating stellar disks.
On a global scale the molecular gas in NGC\,4550 is stable against
gravitational collapse but nevertheless star formation appears to be
ongoing with normal star formation efficiency and gas consumption
time scales. The stellar velocity dispersion in NGC\,4550 resembles 
that of elliptical galaxies. It is therefore likely that a hot 
X-ray emitting plasma limits the lifetime of the molecular gas,
that must arise from a recent ($\ll$1\,Gyr) accretion event.
}

\titlerunning{Molecular gas and dust in NGC\,4550}
\authorrunning{Wiklind \& Henkel}

\maketitle

\section{Introduction}

While counterrotating stellar systems are quite common among
elliptical galaxies, counterrotating stellar {\em disks} are very
rare. Despite dedicated searches (cf. Kuijken et al. 1996), there
are only two known galaxies with large counterrotating stellar
disks; NGC\,4550 (Rubin et al. 1992) and NGC\,7217 (Merrifield \&
Kuijken 1994). There are a few cases which do contain counterrotating 
stellar disk components, but where the secondary component is much 
less extended than the primary one and is confined to the central 
region: NGC\,3593 (Bertola et al. 1996), NGC\,4138 (Jore et al. 
1996) and NGC\,7331 (Prada et al. 1996). On the other hand Bertola 
et al. (1992) and Kuijken et al. (1996) report that $\sim$20--25\% 
of all gas disks found in S0 galaxies counterrotate with respect to 
the stars. If this gas would form stars in such numbers, that more 
than 5\% of the stellar disks were counterrotating, a much higher 
fraction of galaxies with counterrotating stellar components would
be observed.

The most prominent example of a galaxy with two counterrotating
stellar disks is NGC\,4550, an S0/E7 galaxy in the Virgo cluster,
which contains two exponential stellar disks with approximatively
the same scale length and central surface brightness (Rix et al. 
1992). One of the disks also contains a small amount of ionized 
gas (Rubin et al. 1992). The stellar disk corotating with the 
ionized gas has a slightly higher velocity dispersion and lower 
maximum rotational velocity than the stellar disk counterrotating 
with respect to the gas (Rix et al. 1992). Recent results by 
Faundez et al. (2000) also suggest that the disk containing gas 
is sligthly more massive, more vertically extended and dynamically 
hotter than the other. The equivalent widths of the Ca H and K 
features are the same for the two disks and similar to those 
found in elliptical galaxies, suggesting that the stellar disks 
are old and approximately coeval (Rix et al. 1992). NGC\,4550 
contains a low luminosity active galactic nucleus (AGN), seen 
through broad ($\ga$10$^4$\,km\,s$^{-1}$) and double-peaked 
emission lines (Ho et al. 2000).

Is the origin of the massive counterrotating stellar disk in
NGC\,4550 different from that of the numerous less massive 
counterrotating gaseous (and in some cases stellar) disks seen in
other galaxies? Misaligned angular momentum axes are considered to
be clear signatures of merging (cf. Thakar \& Ryden 1996; Thakar
et al. 1997). Strong disk mergers usually lead to the destruction
of the disks and the formation of elliptical systems. Recently,
however, Pfenniger (1999) has shown that similarly sized disks,
but with oppositely aligned angular momentum vectors may merge
without significant heating of the stellar populations. Here 
we report the detection of molecular gas and present optical
images of the dust distribution in NGC\,4550. Consequences for
the origin of this gas and dust are evaluated.

\begin{figure*}
\psfig{figure=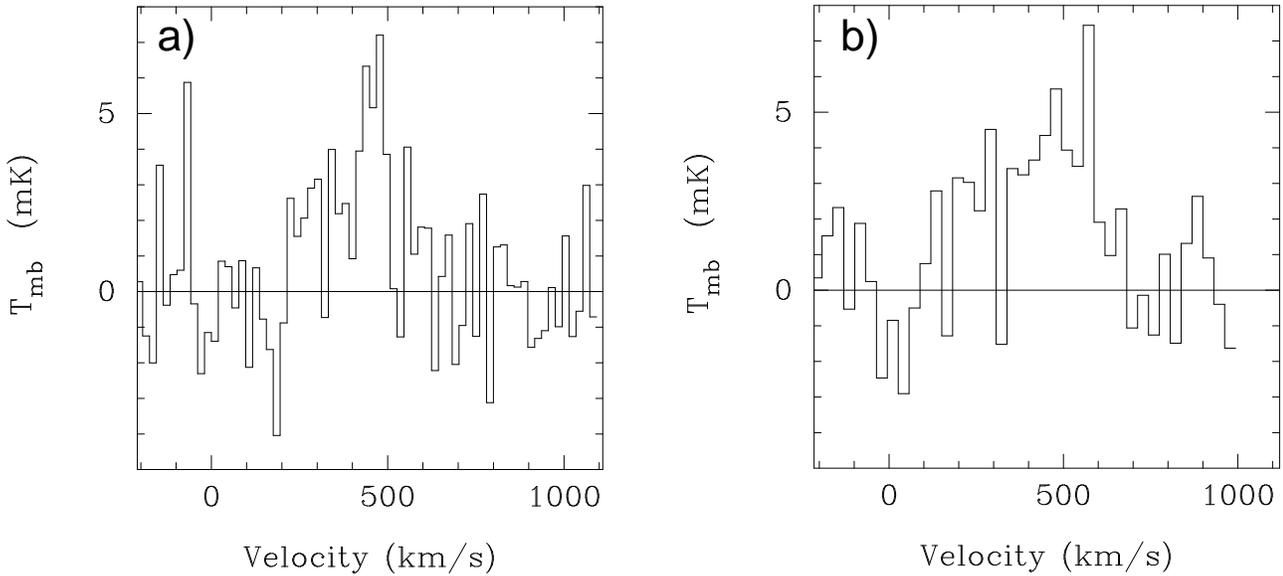,width=17.0cm,angle=-90}
\caption[]{{\bf a)}\ The CO(1--0) emission profile from 
NGC\,4550 obtained with a full width to half power beam size of 
$\theta_{\rm b}$ = 23$''$ in May 2000. The velocity resolution 
is 20\,km\,s$^{-1}$. {\bf b)}\ The CO(1--0) emission obtained 
by convolving the 1992 data with a beamsize $\theta_{\rm b}$ of
= 36$''$ (see text). The velocity resolution is 31\,km\,s$^{-1}$. 
In both cases the telescope beam is centered on the nucleus of 
NGC\,4550, at $\alpha$  = 12$^{\rm h}$\,35$^{\rm m}$\,30.9$^{\rm s}$, 
$\delta$ = $+$12$^{\circ}$\,13$'$\,17$''$ (J2000). The velocity scale 
is heliocentric ($V_{\rm LSR}$ = $V_{\rm HEL}$ + 4.3\,km\,s$^{-1}$)
and the intensity is given in units of main beam brightness 
temperature.}
\label{co}
\end{figure*}

\section{Observations}

\subsection{Molecular data} \label{moldat}

The CO observations were made with the IRAM (Institut de Radio 
Astronomie Millim{\'e}trique) 30-m telescope on Pico Veleta in Spain 
in October 1992 and May 2000. The full width to half power beam size 
was $\theta_{\rm b}$ = 23$''$ (1.9\,kpc at a distance of $D$ = 
16.8\,Mpc; Tully 1988). 3\,mm SiS receivers were used to observe 
the CO(1--0) line with a nutating subreflector, switching symmetrically 
$\pm$240$''$ in azimuth with a frequency of 0.5\,Hz. Typical system 
temperatures were 400\,K (Oct. 1992) and 350\,K (May 2000) on a main 
beam brightness temperature ($T_{\rm mb}$) scale. In 1992, filterbanks 
were used with a channel spacing of 2.6 km\,s$^{-1}$ and a total 
bandwidth of 500 MHz. In 2000, a digital spectrometer was employed; 
the channel spacing was 3.2 km\,s$^{-1}$ and the bandwidth was 
640\,MHz. Final spectra were obtained by adding all individual 
spectra weighted with the inverse of the square of their rms noise, 
removing a first order baseline and binning the channels to an 
effective velocity resolution of 20--31\,km\,s$^{-1}$. 

Typical pointing corrections determined by regular observations of 
nearby continuum sources were 3-4$''$. The temperature scale used 
here is $T_{\rm mb} = T_{\rm A}^{*}\times F_{\rm eff}/B_{\rm eff}$, 
with the forward hemisphere efficiency $F_{\rm eff}$ = 0.95 and the 
main beam efficiency $B_{\rm eff}$ = 0.76. In 1992, image sideband 
rejections were assumed to be 7\,dB, but were not determined for each
individual pointing. An error in the sideband rejection of only 
$\pm$1\,dB would cause a calibration error of $\pm$5\%. In 2000, 
image sideband rejections were $\sim$25\,dB; therefore uncertainties 
in the image sideband rejection do not significantly contribute 
to the total calibration error. As a consequence, the temperature
scale of these measurements is more reliable.

\subsection{Optical data} \label{optdat}

We retrieved Hubble Space Telescope (HST) archival data of NGC\,4550 
obtained during Cycle~4. The data consist of Wide Field/Planetary 
Camera 2 (WFPC2) images with the F555W ($\sim$V--band) and F814W 
($\sim$I--band) filters. The data were obtained on December 10, 1994, 
(proposal 5375, PI: V.C. Rubin), with exposure times of 3$\times$400 
seconds for each filter. The WFPC2 data were processed through the 
standard pipeline where bias, dark, and flatfielding corrections 
were performed and photometry keywords were calculated. Subsequent 
processing was done using standard tasks in IRAF/STSDAS. Multiple 
images were simultaneously co-added and cosmic-rays removed. The 
images shown here are from the Planetary Camera (PC) with a pixel 
scale of 0.046'', corresponding to 3.8\,pc at $D$ = 16.8\,Mpc.

\begin{figure*}
\psfig{figure=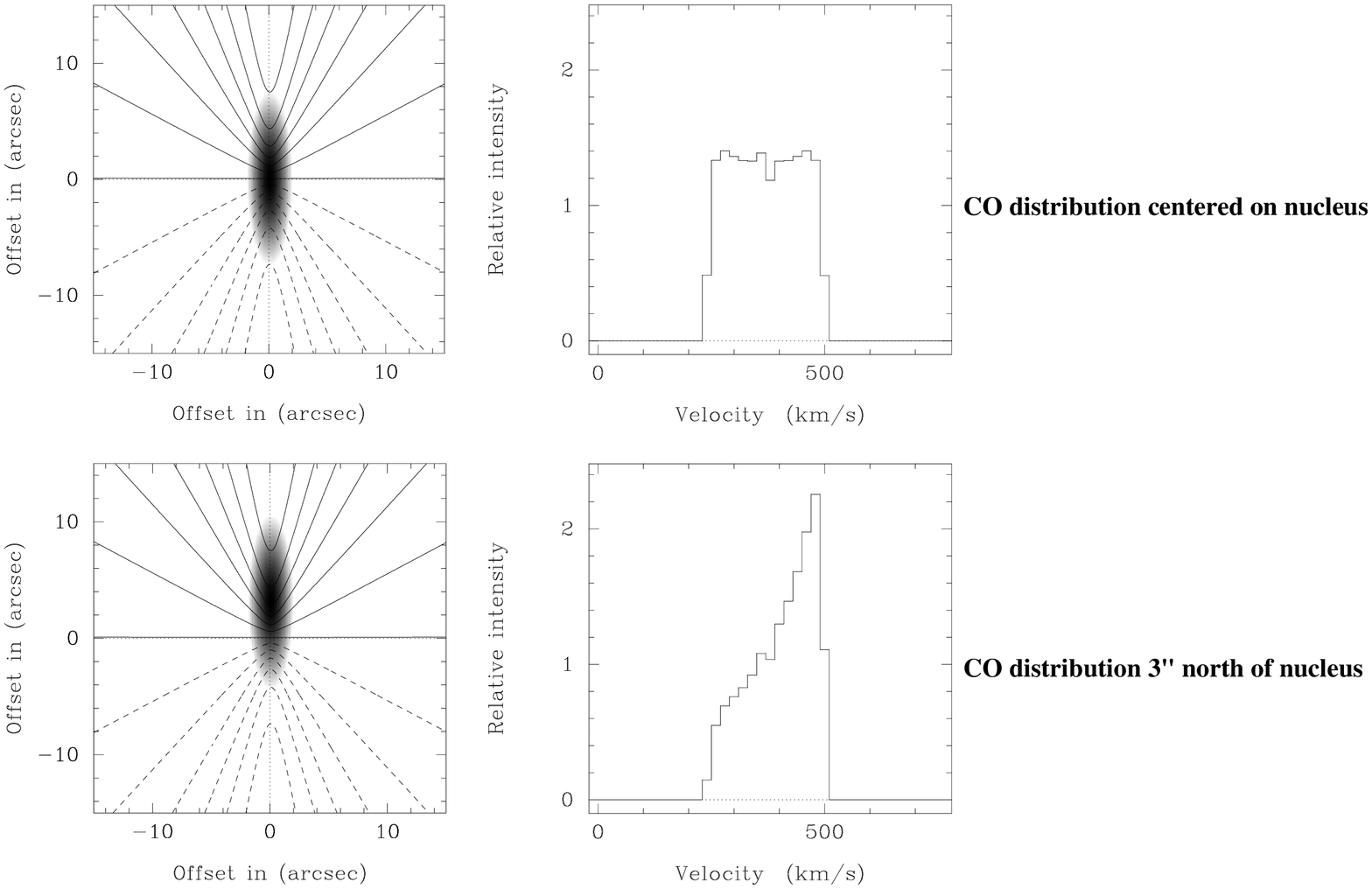,width=18.0cm,angle=-90}
\caption[]{Illustration of the effect of shifting the center 
of a molecular gas distribution relative to the galactic nucleus. 
In this model we `observe' a gaussian shaped CO distribution 
with a full width to half power of $\theta_{\rm s}$ = 14$''$ 
with a beam size of $\theta_{\rm b}$ = 23$''$. The rotation 
curve is taken from Rubin et al. (1992) and the inclination is 
set to 75$^{\circ}$. The left panels show the CO distribution 
in grey scale and the iso-velocity contours in steps of 
20\,km\,s$^{-1}$. The systemic velocity is set to 
380\,km\,s$^{-1}$. The right panels show the resulting CO 
emission profiles. In the top panels we show the result by 
centering the CO distribution on the nucleus. In the bottom 
panels we shift the CO distribution 3$''$ to the north, while 
keeping all other parameters constant. The resulting CO emission 
profile shows a distinct skewness.}
\label{model}
\end{figure*}

\section{Results} \label{res}

\subsection{Molecular gas} \label{molgas}

In Fig.\,\ref{co}a we show the CO(1--0) emission profile observed in
May 2000 towards the center of NGC\,4550. A weak but clear signal is
detected. The emission extends from $\sim$225 to $\sim$535 km\,s$^{-1}$,
which is similar to that of stars and the ionized gas (Rubin et al. 
1992; Rix et al. 1992). The velocity integrated intensity is $0.9 \pm 
0.1$\,K\,km\,s$^{-1}$. With a $N_{\rm H_2}$-to-$I_{\rm CO}$ conversion 
factor of $2.3 \times 10^{20}$\,cm$^{-2}$\,(K\,km\,s$^{-1}$)$^{-1}$, 
and $D$ = 16.8\,Mpc (Tully 1988), this corresponds to an H$_2$ mass 
of $(1.3 \pm 0.1) \times 10^7$\,M$_{\odot}$. The total stellar mass 
can be estimated from the $B$-magnitude, assuming an $M/L$ ratio of 5, 
to be $2 \times 10^{10}$\,M$_{\odot}$. This gives a gas mass fraction 
of only $\sim$0.1\%.

The CO emission profile is skewed relative to the systemic velocity 
of the galaxy ($V_{\rm sys}$ $\sim$ 380\,km\,s$^{-1}$; e.g. Huchtmeier 
\& Richter 1989; Rubin et al. 1997), being stronger at higher velocities. 
Dividing the profile into a low and high velocity part, measuring
the velocity integrated intensities between 220--380 km\,s$^{-1}$
and 380--520 km\,s$^{-1}$ we get
$(I_{high}-I_{low}/I_{rms}) \approx 1.9$.
For a molecular gas distribution being symmetric with respect to the 
minor axis crossing the nucleus, we expect a symmetric emission profile. 
Skewed profiles like the one seen in Fig.~\ref{co}a can then be caused 
by a pointing offset of the telescope. However, in this case we observed 
the same position for several hours, with pointing checks at regular 
intervals. Small pointing uncertainites should therefore result in a 
general broadening of the telescope beam resulting in a symmetric emission 
profile rather than a skewed one. Also, during the course of the 
observations the source rotated with respect to the telescope beam 
counteracting any systematic pointing offset. It is therefore likely 
that the skewed shape of the emission profile in Fig.~\ref{co}a is 
real and reflects an asymmetric distribution of the molecular gas.

In 1992 we observed CO(1--0) emission towards 6 positions in 
the central region of NGC\,4550. Offsets were ($\Delta \alpha$,
$\Delta \delta$) = (0$''$,0$''$), (0$''$,+10$''$), (0$''$,+20$''$), 
(0$''$,--10$''$), (+10$''$,0$''$) and (--10$''$,0$''$), relative
to the center position (see Fig.\,\ref{co}). No emission was 
directly detected above the 3$\sigma$ level in any individual 
position but by convolving the observed data with a telescope 
beam of $\theta_{\rm b}$ = 36$''$, we detect a clear signal.
Omitting the northernmost position from the convolution we get
the CO profile shown in Fig.~\ref{co}b. The exclusion of position
(0$''$,$+$20$''$) does not change the overall appearance significantly.
The emission profile is similar to the one shown in Fig.~\ref{co}a,
i.e. the profile is skewed. The integrated intensity is higher in
Fig.\,\ref{co}b, but in view of the calibration uncertainties (see
Sect.\,\ref{moldat}) and the small number of measured positions
that could be used for the convolution, the temperature scale in
Fig.\,\ref{co}a is more reliable.

In order to test how sensitive the shape of the emission
profile is to offsets of the gas from the nucleus, we made
a simple model of a highly inclined ($i = 75^{\circ}$; cf.
Tully 1988) molecular gas distribution with the major axis
along a north-south direction (for NGC\,4550, $PA$ = 
358$^{\circ}$; Rubin et al. 1992) and the receding part of
the galaxy to the north (Rubin et al. 1992). For the 
molecular gas distribution we assumed a gaussian distribution 
with a full width to half power source size of $\theta_{\rm s}$ 
= 14$''$ (see Sect.\,\ref{dst}). For the velocity field we adopted 
a parametrization of the rotation curve resembling the one 
presented by Rubin et al. (1992). The resulting iso-velocity 
contours, `observed' with a telescope beam size identical 
to the one used in the real observations ($\theta_{\rm b}$ 
= 23$''$), are shown in Fig.\,\ref{model}. The exact shape of the 
rotation curve is not crucial for demonstrating the effect 
from offsetting the center of emission. In the top panels 
we show the case where the gas is centered on the nucleus. 
The observed spectra are symmetric, with a top-hat shape. 
This results from the fact that the gas is sampled mainly 
on the rising part of the rotation curve. In the lower 
panels the gas distribution has been shifted 3$''$ north, 
while keeping all other parameters fixed. The resulting 
emission profile is strongly skewed towards higher velocities 
and resembles the observed CO(1--0) profiles in NGC\,4550
(Fig.\,\ref{co}).

Centering the convolution of the CO profiles observed in
1992 5$''$ north and south of the nuclear position, the 
northern profile shows an integrated intensity that is 
30\%$\pm$10\% larger than the corresponding profile from
the southern position. While this result is only marginally 
significant, it is nevertheless consistent with a skewed 
spatial distribution; most of the CO emission arises from
locations north of the center. 

\subsection{Atomic gas} \label{HI}

There are conflicting reports in the literature regarding
21cm HI observations. While a few early observations reported
detections with integrated fluxes between 1 - 9 Jy\,km\,s$^{-1}$
others, including all recent observations, have resulted in
non-detections. Using the Arecibo antenna DuPrie \& Schneider
(1996) report a non-detection at an rms level of 1.5\,mJy. At
a distance of 16.8 Mpc this corresponds to a 3$\sigma$
upper limit to the HI mass of $7 \times 10^7$\,M$_{\odot}$.
Although this limit is considerably higher than the detected
molecular gas mass, it shows that the total amount of gas in
NGC\,4550 is $< 10^8$\,M$_{\odot}$.

\subsection{Dust} \label{dst}

The asymmetric distribution of the molecular gas indirectly
inferred by the lineshapes of our CO profiles (Fig.\,\ref{co})
can be traced directly by the dust distribution. Dust is visible 
as dark patches in the V-band image and becomes prominent in 
the V$-$~I image (Fig.~\ref{dust}). The dust can be followed 
to a radius of $\sim$7$''$ and is distributed in several arclets, 
reminiscent of spiral arms. The main morphological feature is, 
however, the non-axisymmetric distribution, with the northern 
part being significantly more prominent than the southern part. 
This could in principle be caused by a bar-shaped dust distribution
seen at an angle to the plane of the sky, with the northern side
being closer to the observer. The difference in background light
behind the dust would yield a higher contrast (see Fig.\,3c) in
the north. The extreme difference of the dust obscuration between
northern and southern side of the nucleus appears, however, to be
too large to be explained in this way. An intrinsically
non-axisymmetric distribution is more likely.

NGC\,4550 was detected by IRAS in the 60 and 100$\mu$m bands 
($140 \pm 31$ and $220 \pm 80$ mJy). With $D$ = 16.8\,Mpc the 
total far-infrared (FIR) luminosity amounts to $(1.0 \pm 0.3) 
\times 10^8$\,L$_{\odot}$. The FIR flux corresponds to a dust 
mass
\begin{equation}
M_{\rm dust} = 4.8 \times 10^{-11}
\frac{S_{\nu} D_{\rm Mpc}^{2}}{\kappa_{\nu} B_{\nu}(T_{\rm d})}\ \,
{\rm M}_{\odot},
\end{equation}
where $B_{\nu}$ is the Planck function evaluated for a dust temperature
$T_{\rm d}$ = $39\pm 7$\,K (60/100$\mu$m color temperature),
$D_{\rm Mpc}$ is the distance measured in Mpc (16.8\,Mpc), $S_{\nu}$
is the flux measured in Jansky (here we use the 100$\mu$m flux of
$220 \pm 88$\,mJy) and $\kappa_{\nu}$ is the mass opacity coefficient
for which we used 2.5 m$^2$\,kg$^{-1}$ (Hildebrand 1983) and assumed
$\kappa_{\nu} \propto \nu^{+1}$. The dust mass is $(1.2 \pm 0.9)
\times 10^{5}$ M$_{\odot}$. The large uncertainty is mainly due to
the uncertainty in the dust temperature. The corresponding gas-to-dust 
mass ratio is $\sim 150 \pm 114$, when incorporating a primordial 
abundance of helium. The ionized and atomic gas masses are assumed 
to be negligible in comparison to the molecular gas mass (cf. Rubin 
et al. 1992; DuPrie \& Schneider 1996). Thronson \& Telesco (1986) 
found a typical gas-to-dust mass ratio of $\sim$700 for normal spiral 
galaxies, which reflects the fact that the IRAS bands are not sensitive 
to a dust component colder than $\sim$20\,K. The lower gas-to-dust
ratio found in NGC\,4550 indicates that a relatively warm dust component 
dominates the far infrared continuum.

The heating source of the dust can be (i) the ambient stellar radiation
field, (ii) massive young stars embedded in the dust clouds, (iii) 
photons from a hot X-ray emitting plasma or (iv) collisions with 
electrons from a hot X-ray gas component.

The observed visual magnitude within a 30$''$ aperture centered on 
the nucleus of the galaxy is V=12.31 (Sandage \& Visvanathan 1978). 
This corresponds to an average surface brightness of 19.4 
mag\,arcsec$^{-2}$. Using a solar absolute magnitude $M_{\rm 
V_{\odot}} = 4.83$\,mag we get an average intensity for the ambient 
stellar radiation field of $I_{\ast} = 1.1 \times 10^{-2}$ 
erg\,s$^{-1}$\,cm$^{-2}$\,ster$^{-1}$. The heating rate per dust 
grain by absorption of ambient stellar light is then 
$4\pi I_{\ast}\,Q_{\rm abs}\,\pi a^2 \approx 1.4 
\times 10^{-11}$ erg\,s$^{-1}$, where we have used an absorption 
coefficient of 0.36 and a typical dust grain radius of 0.1$\mu$m 
(cf. Jones \& Merrill 1976; de Jong et al. 1990). The equilibirum 
dust grain temperature can be estimated by equating the heating 
and cooling rates and becomes $\sim$25\,K. This is 2$\sigma$ lower 
than the observed $39 \pm 7$ K and suggests an additional heating
source for the dust. Although there are no obvious signs for 
massive stars in NGC\,4550 (cf. Fig.~\ref{dust}), the gas mass 
derived from the CO emission and from the FIR luminosity are 
consistent with each other assuming a normal gas-to-dust mass 
ratio. Hence it cannot be excluded that young massive stars are
embedded in the molecular gas. In Sect.\,\ref{xray} we will see 
that X-ray heating, either through X-ray photons or hot electrons, 
is inefficient in comparison with heating by the ambient stellar 
radiation field.

\subsection{X-ray properties}\label{xray}

NGC\,4550 is undetected at X-rays. Einstein data give $L_{\rm X} < 
4.1 \times 10^{40}$\,ergs\,s$^{-1}$ (Fabbiano et al. 1992). It 
is nevertheless likely that a hot X-ray emitting plasma is present
at some level. Stellar mass loss from evolved stars will contribute
$\sim 0.15 \,(L_{\rm B}/10^{10}\,L_{\odot})$ M$_{\odot}$\,yr$^{-1}$
of gas to the interstellar medium (cf. Faber \& Gallagher 1976; 
Sarazin 1990). The large stellar velocity dispersion in NGC\,4550, 
caused by the counterrotation, will heat this gas to X-ray temperatures.
The gas lost from evolved stars will have an energy per unit mass of
$3kT/2\mu m_p$, similar to the energy per unit mass of stars, which is
of the order $\sigma_{\ast}^2/2$ (using the one-dimensional velocity
dispersion since stellar orbits are confined to a plane). After 
thermalization this leads to an equilibrium temperature of (cf. 
Sarazin 1990)
\begin{equation}
T_{\ast} = \frac{\mu m_p\sigma_{\ast}^2}{3\:k}
\approx 2 \times 10^6
\left(\frac{\sigma_{\ast}}{300\,{\rm km}\:{\rm s}^{-1}}\right)^2\ {\rm K}\ \ .
\end{equation}
An additional source of heating comes from supernovae, but due to 
the high relative velocity dispersion $\sigma_{\ast}$ in NGC\,4550, 
it is likely that energy input from stellar mass loss will dominate 
the heating. Assuming that the gas remains gravitationally bound 
to the galaxy and cools radiatively, the resulting X-ray luminosity
can be approximated as 
\begin{eqnarray}
L_{\rm X} & \approx & \frac{1}{2}\,\sigma_{\ast}^{2}\,
            \left[\frac{\dot{M}_{\ast}}{L_{\rm B}}\right]
	    \,L_{\rm B} \nonumber \\ 
          & \approx & 4.3 \times 10^{39}
            \left(\frac{\sigma_{\ast}}{300\,{\rm km/s}}\right)^{2}
	    \:\left(\frac{L_{\rm B}}{10^{10}\:{\rm L_{\odot}}}\right)\ 
	    {\rm ergs}\:{\rm s}^{-1}\ ,
\end{eqnarray}
where the term in bracketts is the stellar mass loss rate per blue
luminosity (cf. Faber \& Gallagher 1976; Sarazin 1990). For NGC\,4550,
with $\sigma_{\ast} \approx 300$ km s$^{-1}$ and $L_{\rm B} = 0.3 
\times 10^{10}$ L$_{\odot}$, we get $L_{\rm X} \approx 1 \times 
10^{39}$ ergs s$^{-1}$. This is consistent with the existing upper 
limit for NGC\,4550 of $4 \times 10^{40}$ ergs s$^{-1}$ (Fabbiano 
et al. 1992). The time scale for accumulating an X-ray gas mass 
equivalent to the molecular gas mass is
\begin{equation}
\Delta t = \left[\frac{\dot{M}_{\ast}}{L_{\rm B}}\right]^{-1}
L_{\rm B}^{-1} M_{\rm gas}\ \approx 300\ {\rm Myr}\ .
\end{equation}
$[\dot{M}_{\ast}/L_{\rm B}]$ is the stellar mass loss rate
from evolved stars, assumed to be $1.5 \times 10^{-11}$ 
M$_{\odot}$\,year$^{-1}$\,L$_{\odot}^{-1}$ (cf. Faber \& 
Gallagher 1976; Sarazin 1990).

Hence, a substantial hot X-ray emitting plasma can be established
in the center of NGC\,4550 on a relatively short time scale. This
plasma can contribute to the heating of the dust component as well
as contribute to its destruction through sputtering.

Assuming a core radius for the X-ray gas of 1\,kpc (12$''$ at $D$ =
16.8\,Mpc), the average X-ray intensity in the central region of 
NGC\,4550 is $I_{\rm X} = 3 \times 10^{-6}$ ergs s$^{-1}$ cm$^{-2}$ 
ster$^{-1}$, resulting in a heating rate per dust grain of
$4\pi\:I_{\rm X}\:\pi\:a^2 \approx 1 \times 10^{-14}$ ergs s$^{-1}$.
This is three orders of magnitude smaller than the heating rate from
the ambient stellar light (Sect.\,\ref{dst}). The X-ray photon 
contribution to the heating of the dust can thus be neglected.

The hot electrons in the core of the X-ray emitting gas will collide
with dust grains. The flux of hot electrons is $F_{e} = n_e (8 k 
T_e/\pi m_e)^{1/2} = 6.2 \times 10^8 n_e (T_e/10^6\,K)^{1/2}$ 
cm$^{-2}$ s$^{-1}$ (cf. de Jong et al. 1990). Most of the kinetic 
energy of the hot electrons will be deposited in the grains upon 
collision, resulting in a heating rate $(3/2)\:F_e\:k\:T_e\:\pi 
a^2 \approx 4.0 \times 10^{-11}\:n_e\:(T_e/10^6 K)$ ergs\,s$^{-1}$. 
Assuming an electron density of $10^{-2}$ cm$^{-3}$ and the above 
derived gas temperature of $2 \times 10^6$ K, the heating rate 
per dust grain due to hot electrons becomes $1 \times 10^{-12}$
ergs\,s$^{-1}$. This is much larger than the heating rate due to 
X-ray photons but still ten times less than the heating rate due 
to the ambient stellar radiation field.

The hot plasma will destroy the dust grains through `sputtering' (cf.
Draine \& Salpeter 1979). For diffuse gas (i.e. $A_{\rm V} \leq 1$, the
time scale is very short: $\Delta t \approx 2 
\times 10^{5}\,(n_e/{\rm cm}^{-3})^{-1}\,(a/0.1\mu m)$\,yr
(Draine \& Salpter 1979). With $n_e \approx 10^{-2}$ cm$^{-3}$ and
a typical grain size of 0.1$\mu$m, $\Delta t$ is only 20 Myr.
After $\sim$100 Myr the only surviving gas clouds are those which
are optically opaque with $A_{\rm V} > 3$ (cf. de Jong et al. 1990).
The lifetime of these opaque clouds depends strongly on geometry 
and physical conditions. De Jong et al. (1990) estimated typical 
parameters for this dense gas component in NGC\,4696, the central
elliptical galaxy of the Centaurus cluster, and found values similar 
to the cores of molecular clouds in our Galaxy. Star formation 
could be sustained in these environments.

To summarize: Mass lost from evolved stars will be heated to a 
temperature of $\sim 2 \times 10^6$ K on a relatively short time scale 
due to the high (one dimensional) stellar velocity dispersion in 
NGC\,4550. The X-ray luminosity is consistent with the upper limit 
from Einstein data. The X-ray gas will not contribute significantly 
to the heating of the dust grains but will destroy any diffuse 
component through sputtering on a time scale of 20--100 Myr. The 
only remaining dusty molecular gas component is associated with
dense clouds of high optical opacity, possibly sustaining star formation.

\subsection{Star formation} \label{starformation}

Assuming for the moment that the FIR luminosity is powered entirely by
young and massive stars and that the initial mass function (IMF) is of
Salpeter type with cut-offs at 0.1 and 100 M$_{\odot}$ (cf. Thronson \&
Telesco 1986), the star formation rate (SFR) becomes 0.065\,M$_{\odot}$\,yr$^{-1}$.
The ratio of $L_{\rm FIR}/M_{\rm H_2}$ is usually taken as a measure
of the star formation efficiency. In NGC\,4550 this ratio is $8 \pm 
2$ L$_{\odot}$/M$_{\odot}$, which is similar to the values found
for normal spiral galaxies. The gas consumption time scale is defined as
the ratio $M_{\rm H_2}$/SFR, which for NGC\,4550 is $2 \times 10^{8}$ years.

Is star formation really going on in the center of NGC\,4550? Kennicutt 
(1989) suggested that significant star formation only occurs only when 
the gas surface density $\Sigma_{\rm gas}$ exceeds a critical value 
$\Sigma_{\rm gas}^{\rm crit}$. The latter value depends on the galaxy 
rotation and on the gas velocity dispersion. The rotation curve presented 
by Rubin et al. (1992, 1997) shows that the ionized gas rotates approximately 
like a solid body in the inner 7$''$, with a maximum rotational velocity 
of 160\,km\,s$^{-1}$. Within the region of solid body rotation we can 
express the critical gas surface density as
\begin{equation}
\Sigma_{\rm gas}^{\rm crit} = 0.138 \alpha \Omega c\ \ {\rm M}_{\odot}
{\rm pc}^{-2},
\end{equation}
where $\Omega$ is the angular velocity in units of km\,s$^{-1}$\,kpc$^{-1}$,
$c$ is the gas velocity dispersion in units of km\,s$^{-1}$ and $\alpha$
is a dimensionless constant of order unity. For solid body rotation,
this relation is independent of galactocentric distance as long as
the velocity dispersion remains constant. We do not know the extent 
of the molecular gas disk nor its velocity dispersion, but reasonable 
assumptions are that it coexists with the dust (Fig.~\ref{dust}), 
having a radial extent of $\sim$7$''$, and that it has a velocity 
dispersion  $\sim$10\,km\,s$^{-1}$. Furthermore, assuming that the 
molecular gas corotates with the ionized gas, the critical gas surface 
density is $\Sigma_{\rm gas}^{\rm crit} \approx 260 
(c/10$km\,s$^{-1})$\,M$_{\odot}$\,pc$^{-2}$. The observed gas surface 
density, assuming a homogeneous disk with a radial extent of 7$''$ 
(0.58\,kpc) is $\Sigma_{\rm gas} \approx 17$\,M$_{\odot}$\,pc$^{-2}$,
where we have included He at primordial abundances. The ratio 
$\Sigma_{\rm gas}/\Sigma_{\rm gas}^{\rm crit} = 0.07 (c/10$km\,s$^{-1})^{-1}$.
Since we found the molecular gas to be distributed non-axisymmetrically
in the same sense as the dust seen in Fig.~\ref{dust}, the actual gas 
surface gas density is likely a factor of two larger. Nevertheless, 
even if the velocity dispersion is only 5\,km\,s$^{-1}$, the gas 
surface density is only $\sim 0.25 \Sigma_{\rm gas}^{\rm crit}$.
This means that the molecular gas is likely to be stable against 
gravitational collapse on a global scale. Smaller scales with potentially 
higher gas surface densities will be discussed in the following section.

\begin{figure*}
\caption[]{HST WFPC2 images of the central region of NGC\,4550. Only the PC chip
is shown, with a pixel size of 0.046$''$. Three images are displayed, centered
on the nucleus: F555W (V--band), F814W (I-band) and the V-I image (inverted).
North is up and east to the left.}
\label{dust}
\end{figure*}

\section{Discussion}\label{discussion}

Although NGC\,4550 is exceptional in that it contains two similar stellar
disks counterrotating with respect to each other, there are few if any
morphological and photometrical properties that distinguish it from other
early-type galaxies. Surface photometry of 14 E and S0 galaxies in the 
Virgo cluster, obtained with the Hubble Space Telescope (HST) and including 
NGC\,4550 (Jaffe et al. 1994) reveals nuclear disks in all of the early 
type galaxies later than E4. A nuclear dust component was also seen in most 
of the early type galaxies. Jaffe et al. (1994) did not find any property 
distinguishing NGC\,4550 from the other 13 galaxies, in spite of the 
counterrotating stellar disks. Furthermore, the FIR properties of the early 
type galaxies in the Jaffe et al. (1994) sample are similar to those of 
NGC\,4550. Hence, the presence of $\sim 10^7$\,M$_{\odot}$ of molecular 
gas in the center of NGC\,4550 is not exceptional for its morphology 
(see also van Dokkum \& Franx 1995). The only peculiarity which may point 
to the presence of the counterrotating disks is the non-axisymmetric 
distribution of the molecular gas and dust in the center of NGC\,4550.

Several numerical/theoretical studies of the dynamics of disks containing
counterrotating stars and/or gas has shown that the fastest growing unstable
mode is an $m=1$ instability (Merrifield \& Kuijken 1994; Lovelace et al.
1997; Thakar et al. 1996; Thakar \& Ryden 1998; Garc\'{\i}a-Burillo et al. 
2000).  This applies to the cases where the mass fractions of the pro- 
and retrograde mass components are of comparable size. The result is a 
tightly wound one-armed spiral structure within a few hundred pc from 
the nucleus. Initially the $m=1$ instability can also be manifest through 
a lopsidedness of the mass distribution. This latter feature is, however, 
relatively short-lived, 0--500 Myr. On time scales longer than 1 Gyr an 
$m=2$ instability dominates (cf. Garc\'{\i}a-Burillo et al. 2000). This 
takes the form of a bar or oval configuration.

An extensive study of instabilities in counterrotating self-gravitating
stellar disks was presented by Sellwood \& Merritt (1994). They found
that for kinematically cool systems, lopsidedness is the only instability
which is unique for counterrotating stellar systems (this applies for
isolated galaxies). Contrary to the studies mentioned above, Sellwood
\& Merritt found the lopsidedness to persist for long periods and suggested
that this feature could be used to identify galaxies with counterrotating
stellar subsystems.

The dust distribution in NGC\,4550 is reminiscent of a tightly wound
spiral arm structure, extending to a galactocentric distance of
$\sim$600\,pc, but with a pronounced lopsidedness.  
As described in Sect.\,\ref{res}, the CO emission is consistent with an
asymmetric distribution of the molecular gas, in the same manner as the
distribution of the dust component. The kinematics of the CO emission
suggests that molecular gas is present on both sides of the nucleus, but
that the northern side is dominant. In view of the above mentioned 
numerical simulations, the most likely explanation for the lopsided gas 
and dust distribution in NGC\,4550 is an instability caused by the 
counterrotating disks. As pointed out by Sellwood \& Merritt (1994), 
the lopsided instability is strongest for the coolest systems, i.e. 
the gas. It remains to be shown whether these features can be longlived. 
If not, the lopsidedness implies a recent accretion event ($\leq$0.5 Gyr).

In Sect.\,\ref{xray} we argued for the existence of a hot X-ray
emitting plasma in the central region of NGC\,4550. Although this
X-ray component was found to give a neglible contribution to the
heating rate of the dust grains, it limits the lifetime of
any diffuse dust component to less than $\sim$100 Myrs. A dense
gas component can survive for longer time scales. The physical
characteristics needed for molecular gas surviving in a hot X-ray
emitting plasma implies small and dense molecular clouds, similar
to molecular cloud cores in in our own Galaxy. These cores could
very well sustain star formation despite that the global surface
density of molecular gas in NGC\,4550 was shown to be too low to
be unstable to gravitational collapse (Sect\,\ref{starformation}).
This star formation activity would contribute the extra heating
suggested to explain the high dust temperature of $39 \pm 7$ K
(see Sect.\,\ref{dst}).

The only other known system with a significant counterrotating
stellar disk and molecular gas is NGC\,3593. In contrast to
NGC\,4550, this edge-on S0/a galaxy contains two stellar disks
with different central surface brightness and different scale
heights (Bertola et al. 1996). The smaller disk, which makes up
$\sim$18\% of the total disk stellar mass, also contains
significant amounts of ionized, atomic as well as molecular gas
(Wiklind \& Henkel 1992; Corsini et al. 1998; Garc\'{\i}a-Burillo et 
al. 2000). The atomic and molecular gas masses are $1.8 \times 10^8$
and $4.5 \times 10^8$\,M$_{\odot}$, respectively (Krumm \& Salpeter
1979; Wiklind \& Henkel 1992). Including the counterrotating gas
components, the mass ratio of the two disks is $\sim$0.3, with
the gas making up $\sim$40\% of the total mass of the counterrotating
disk ($M_{\rm stars} + M_{\rm gas}$). This is in stark contrast to
NGC\,4550, where the mass ratio of the two disks is $\sim$1 with
the gas making up $\sim$0.1\% of the mass of one of the disks.

NGC\,3593 shows several signs of vigorous star formation activity
(cf. Henkel \& Wiklind 1997). It has a relatively large FIR luminosity
suggesting an SFR of 1.6\,M$_{\odot}$\,yr$^{-1}$ (Wiklind \& Henkel 1992),
25 times higher than NGC\,4550. The star formation activity is confined 
to a central ring with a radius of $\sim$300\,pc. The ring is seen in 
both ionized and molecular gas (Wiklind \& Henkel 1992; Corsini et al. 
1998; Garc\'{\i}a-Burillo et al. 2000). The atomic gas has only been 
observed with a single dish telescope but appears to be significantly 
more extended than both the ionized and molecular gas component. It is 
thus possible that in NGC\,3593 we are witnessing the transformation of 
accreted gas, with an angular momentum opposite to that of the main 
stellar disk, into stars. Once the gas is consumed, NGC\,3593 will 
look like an S0 galaxy with a counterrotating stellar population 
comprising $\sim$25\% of the total stellar disk mass. Although less 
than in NGC\,4550, it is still significant in comparison to the limits 
obtained for counterrotating stellar components in the sample of 28 
S0 galaxies by Kuijken et al. (1996).

In the case of NGC\,4550 there is only weak and indirect evidence for 
on-going star formation. Regardless of whether star formation is
taking place or not, the age of the observed molecular gas and dust
must be low. Since mass loss from evolved stars would provide cool
gas associated with both stellar disks that are old and approximately 
coeval (see Rix et al. 1992), the gas (and the dust) must be the result 
of a relatively recent ($\ll$1\,Gyr) accretion event not related to the 
main build-up of the two counterrotating stellar populations. Searching
for potential sources for a mass transfer, the nearby elliptical galaxy 
NGC\,4551 ($V_{\rm sys}$ $\sim$ 1170\,km\,s$^{-1}$ versus 380\,km\,s$^{-1}$ 
for NGC\,4550) is a highly unlikely target; the accretion of a dwarf galaxy
is therefore a more reasonable scenario.

\acknowledgements
TW acknowledges support from NFR (the Swedish Natural Science Research
Council) for a research grant.


\end{document}